\documentclass[fleqn,usenatbib]{mnras}

\usepackage[T1]{fontenc}

\DeclareRobustCommand{\VAN}[3]{#2}
\let\VANthebibliography\thebibliography
\def\thebibliography{\DeclareRobustCommand{\VAN}[3]{##3}\VANthebibliography}

\usepackage{graphicx}	
\usepackage{amsmath}	
\usepackage{amssymb}	

\usepackage{comment}    
\usepackage{float}

\newcommand{\angstrom}{\textup{\AA}}

\title[Empirical abundance calibration]{Chemical abundances in Seyfert galaxies -- VI. Empirical abundance calibration}
\author[O. L. Dors]{
Oli L. Dors$^{1}$\thanks{E-mail: olidors@univap.br}
\\
$^{1}$Universidade do Vale do Para\'iba, Av. Shishima Hifumi, 2911, Cep
12244-000, S\~ao Jos\'e dos Campos, SP, Brazil \\
}

\date{Accepted XXX. Received YYY; in original form ZZZ}

\pubyear{2015}

\begin{document}
\label{firstpage}
\pagerange{\pageref{firstpage}--\pageref{lastpage}}
\maketitle

\begin{abstract}
We derived a bi-dimensional calibration between  the emission line ratios
 $R_{23}= ([\ion{O}{ii}]\lambda3726 +\lambda3729 +[\ion{O}{iii}]\lambda4959 + \lambda5007 )/\rm H\beta$,
${\rm P}=[([\ion{O}{iii}]\lambda4959+\lambda5007)/{\rm H}\beta]/R_{23}$
and the oxygen abundance relative to hydrogen (O/H)
in the gas phase of Seyferts~1 and 2 nuclei. In view of this, emission-line intensity ratios for a sample of objects taken from the Sloan Digital Sky Survey Data Release 7 (SDSS-DR7) measured by the  MPA/JHU group and
direct estimates of O/H based on $T_{\rm e}$-method, adapted for AGNs, are considered.
 We find no variation of $R_{23}$  observed along the radii of AGNs which shows that this line ratio is a good oxygen abundance (O/H) indicator for the class of objects considered in this work.
The derived O/H = f($R_{23}$, P) relation produces O/H values similar to estimations via $T_{\rm e}$-method in a wide range of
metallicities ($\rm 8.0 \: \lesssim \: 12+\log(O/H) \: \lesssim \: 9.2$). Conversely to star-forming regions
in the high metallicity regime, $R_{23}$ shows a positive correlation trend with O/H in AGNs. This indicates that the hardness of ionizing  radiation is not affected by the metallicities in these objects or  Narrow Line Regions (NLRs) 
are not significantly  modified by changes in the Spectral Energy Distribution due to metallicity variations.

\end{abstract}

\begin{keywords}
galaxies: Seyfert -- galaxies: active -- galaxies: abundances --ISM: abundances
--galaxies: evolution --galaxies: nuclei 
\end{keywords}



\section{Introduction}

Active Galactic Nuclei (AGNs) and Star-forming regions (SFs) present in their spectra
strong metal and hydrogen emission lines, whose relative intensities can be used
to derive the metallicity and/or the abundances of heavy elements.  These features make these objects essential for studying the chemical evolution of galaxies.

It is widely known that the most dependable approach for determining the chemical abundance of heavy elements (e.g., O, N, S) in the gas phase of SFs and Planetary Nebulae is primarily based on direct measurements of the electron temperature ($T_{\rm e}$), which is commonly referred to as the $T_{\rm e}$-method  (for a review, see \citealt{2017PASP..129h2001P, 2017PASP..129d3001P, 2019A&ARv..27....3M}).  Basically, this method consists of determining the  
 $T_{\rm e}$  of the gas phase through emission-line intensity ratios
emitted by a given ion  and originated  in transitions from two
levels with considerable different excitation energies.  For example, the $T_{\rm e}$
for the gas regions where the $\rm O^{+}$, $\rm O^{2+}$ and  $\rm N^{+}$ ions
are located
can be determined through  the [\ion{O}{ii}]($\lambda$3726+$\lambda$3729)/($\lambda$7319+$\lambda$7330), [\ion{O}{iii}]($\lambda$4959+$\lambda$5007)/$\lambda$4363 and 
[\ion{N}{ii}]($\lambda$6548+$\lambda$6584)/($\lambda$5755) line ratios, respectively (e.g., \citealt{2008MNRAS.383..209H}). However, for
most extragalactic  objects where the $T_{\rm e}$-method can be applied, which has only the [\ion{O}{iii}]($\lambda$4959+$\lambda$5007)/$\lambda$4363 line ratio measured, it is only possible to estimate
the temperature for the high ionization zone (e.g., \citealt{1998AJ....116.2805V}). In these cases, temperatures for
the low ionization zones  are obtained from empirical (e.g., \citealt{2003ApJ...591..801K, 2006MNRAS.370.1928P, 2007MNRAS.375..685P, 2009ApJ...700..654E, 2015ApJ...806...16B, 2016ApJ...830....4C, 2020A&A...634A.107Y}) or from photoionization models (e.g., \citealt{1986MNRAS.223..811C, 1992MNRAS.255..325P, 1992AJ....103.1330G, 1997ApJS..108....1I, 2000MNRAS.311..329D, 2009MNRAS.398..949P, 2014MNRAS.441.2663P}) relations, which can introduce
some uncertainty in the abundance determinations
(e.g., \citealt{2020MNRAS.497..672A}).

The  reliability of the $T_{\rm e}$-method is  supported  by the agreement between
oxygen abundance relative to hydrogen (O/H) estimates  in \ion{H}{ii}  regions located in the  solar neighborhood and those derived through  the weak interstellar \ion{O}{i}$\lambda$1356 {\angstrom} line towards the stars (see \citealt{2003A&A...399.1003P} and references therein). Moreover, a consonance has been found between oxygen abundances obtained for SFs and B-type stars in the Milky Way and other nearby galaxies (see \citealt{toribio17} and references therein).
However, determination of
$T_{\rm e}$ requires measurements of  auroral emission lines (e.g., [\ion{O}{iii}]$\lambda4363$ {\angstrom},
 [\ion{N}{ii}]$\lambda$5755 {\angstrom},  [\ion{S}{iii}]$\lambda6312$ {\angstrom}), which are generally weak (about
 100 times weaker than H$\beta$) in the spectrum of objects with high metallicity
 and/or low excitation (e.g., \citealt{1998AJ....116.2805V, 2007MNRAS.382..251D, 2008A&A...482...59D}). To circumvent this limitation  of the $T_{\rm e}$-method,
 \citet{10.1093/mnras/189.1.95}, following the original idea
 of \citet{1976ApJ...209..748J},  proposed the use of the
 $R_{23}$=([\ion{O}{ii}]$\lambda3726+\lambda3729$+[\ion{O}{iii}]$\lambda$4959+$\lambda$5007)/H$\beta$ line ratio as O/H abundance indicator for SFs. After the aforementioned pioneering work, several authors have proposed
 calibrations for SFs between $R_{23}$ and the O/H abundance 
 as well as considerations for other line ratios (for a review see \citealt{lopez2010massive}). Basically, there are two ways to obtain a calibration between strong emission line ratios and O/H (a metallicity tracer), i.e., assuming predictions from photoionization models (e.g., \citealt{1991ApJ...380..140M, 2002ApJS..142...35K}) and by using observational emission lines and O/H abundance values derived from $T_{\rm e}$-method (e.g., \citealt{1994ApJ...429..572S, 2000A&A...362..325P,  2001A&A...369..594P, 2007A&A...462..535Y, 2013A&A...559A.114M, 2019ApJ...872..145J}). 
 This method of utilizing calibration to estimate elemental abundances is known as strong-line method.
It has been established that, for SFs,  the majority of the strong-line methods based on  theoretical models overestimate the O/H abundance as compared  to the results obtained from the $T_{\rm e}$-method (e.g., \citealt{2007A&A...462..535Y}), where
the discrepancies are in order of 0.2-0.3 dex  \citep{lopez2010massive}.

In comparison with SFs, there are few abundance estimates for AGNs derived from the $T_{\rm e}$-method in the literature. In fact, it appears that  most complete abundance determinations in AGNs
based on $T_{\rm e}$-method were carried out by \citet{1975ApJ...197..535O}, who derived
the He, O, N, Ne, and Fe abundances, in relation to the hydrogen,  in the gas phase of 3C\,405 (Cygnus~A). The majority of the other studies (e.g., \citealt{dors2015central, 2020MNRAS.496.3209D, 2008ApJ...687..133I}) have been focused on determining only the O/H abundance and a few for the N/H (e.g., \citealt{2020MNRAS.496.2191F}). Furthermore, for the strong-line method based on narrow optical lines of AGNs, there are only theoretical calibrations proposed by \citet{1998AJ....115..909S}  as well as semi-empirical calibrations proposed by \citet{2017MNRAS.467.1507C}, \citet{2020MNRAS.492.5675C} and \citet{2021MNRAS.501.1370D}. These calibrations are primarily based on photoionization models, which may have some uncertainties. Firstly, the foregoing AGN calibrations consider lines emitted by oxygen and nitrogen, so it is important to assume a correct relation between O and N in the models \citep{2009MNRAS.398..949P}. However, the N/H abundance is barely known in AGNs. In fact, 
 \cite{dors17}, who used detailed photoionization models to reproduce narrow optical
 ($3000 \: < \: \lambda(\text{ \AA}) \: < \:7000$) emission lines
 of a sample of AGNs, presented the first quantitative nitrogen abundance determination for a small sample of 44 Seyfert 2 nuclei in the local universe ($z \: \la \: 0.1$; see also \citealt{2002ApJ...564..592H, contini2017abundance, 2019MNRAS.489.2652P, 2020MNRAS.496.2191F}). Secondly, photoionization
models are subject to intrinsic uncertainties,  e.g., all relevant physical process are not treated
correctly, inaccurate atomic data use, spherical geometry consideration, etc (see \citealt{1984PASP...96..593N, 2002RMxAC..12..219V, 2003ApJ...591..801K}).
 
Recently, \citet{2020MNRAS.496.3209D} investigated the discrepancy between O/H abundance  estimations for narrow-line regions (NLRs) of
 Seyferts~2  derived by using $T_{\rm e}$-method and 
those derived from photoionization models. These authors found that 
the derived discrepancies are mainly due to the inappropriate
use of the relations between temperatures of the low ($t_{2}$) and high ($t_{3}$) ionization  gas zones derived for \ion{H}{ii} regions in AGN chemical abundance studies. 
In addition, \citet{2020MNRAS.496.3209D}, using a photoionization model grid,  derived a new expression  for   
the  $t_{2}$-$t_{3}$ relation  valid for Seyfert 2 nuclei which reduces the O/H discrepancies
between the abundances obtained from strong-line methods and those derived from $T_{\rm e}$-method by $\sim$0.4 dex. This new methodology, combined with the very large sample of spectroscopic
data made available by the  Sloan Digital Sky Survey (SDSS, \citealt{2000AJ....120.1579Y}),
will help to build an empirical calibration for AGNs, which is not available in the literature reviewed thus far. 

Following from above, the emission-line 
intensities of the SDSS-DR7 \citep{2009ApJS..182..543A} measured 
by the MPA-JHU group\footnote{Max-Planck-Institute for Astrophysics and John Hopkins University}
and the methodology proposed by \citet{2020MNRAS.496.3209D}
are used in this study to calculate the O/H abundance for a sample of Seyfert~1 and 2 nuclei.
Thereafter, following the methodology considered by \citet{2000A&A...362..325P, 2001A&A...369..594P} and the same supposition for SFs -- 
 the strong oxygen lines 
[\ion{O}{ii}]$\lambda\lambda$3726 {\angstrom}, 3729 {\angstrom} and [\ion{O}{iii}]$\lambda\lambda$4959 {\angstrom}, 5007{\angstrom}
contain the necessary information required to accurately derive the O/H abundance \citep{1991ApJ...380..140M} --
we obtain a bi-dimensional empirical calibration between the  
$R_{23}$ and P=([\ion{O}{iii}]$\lambda$4959 + $\lambda$5007/H$\beta$)/$R_{23}$
line ratios and O/H abundance, valid for AGNs. 
The present study is organized as follows.   In Section~\ref{oxy}, 
the  observational data and the methodology
used to estimate the oxygen abundance are presented.
The resulting calibration and 
the discussion are presented in Sects.~\ref{res} and \ref{disc}, respectively. Finally, the conclusion of the outcome is given in Sect.~\ref{conc}.

\section{methodology}
\label{oxy}
To obtain a calibration between strong oxygen emission lines and the 
O/H abundance for AGNs, we selected from the SDSS DR7 spectroscopic data of confirmed
sample of  types 1 and 2 Seyfert nuclei. These data were used to calculate the O/H abundance through the $T_{\rm e}$-method and, afterwards, an empirical calibration was derived. In what follows, each one of the procedures mentioned above is described.

\subsection{Observational data} 
\label{observ}

We used optical emission-line intensities of Seyferts~1 and 2 nuclei taken from 
the Sloan Digital Sky Survey (SDSS, \citealt{2009ApJS..182..543A}) DR7   and
 presented in \citet{2020MNRAS.492..468D} (hereafter Paper I). 
 From these data, we considered only the intensities
 (in relation to H$\beta$) of the emission lines
[\ion{O}{ii}]$\lambda$3726 {\angstrom}+$\lambda$3729 {\angstrom},  
[\ion{O}{iii}]$\lambda$4363 {\angstrom}, 
[\ion{O}{iii}]$\lambda$5007 {\angstrom}, 
H$\alpha$, 
[\ion{N}{ii}]$\lambda$6584 {\angstrom}, 
[\ion{S}{ii}]$\lambda$6716 {\angstrom} and
[\ion{S}{ii}]$\lambda$6731 {\angstrom}.
The lines measurements were carried out by the MPA/JHU group and they 
are reddening corrected (see Paper~I). 

The  Seyferts~1 and 2 classifications were obtained
by cross-correlation between the identification of each object in the
SDSS data and   in a catalogue provided by  
NED/IPAC\footnote{ned.ipac.caltech.edu}
(NASA/IPAC Extragalactic Database). Here, we only considered objects for which the [\ion{O}{iii}]$\lambda$4363 {\angstrom} line has a measurement error of less than 50\,\% of its intensity. Thereafter, to minimize the contribution of  SF emission  to the observed AGN fluxes, we only considered objects which
are more than 0.1 dex (see \citealt{2006MNRAS.372..961K}) 
above the demarcation lines 
proposed by \citet{2001ApJ...556..121K} to separate AGNs from SFs, where
objects with
\begin {equation}
\label{eq1}
\mathrm{\log([\ion{O}{iii}]\lambda5007/H\beta}) >  \frac{0.61}{\mathrm{\log([\ion{N}{ii}]\lambda6584/H\alpha)-0.47}} + 1.19
\end {equation}
and
\begin{equation} 
\label{eq2}
\mathrm{\log([\ion{O}{iii}]\lambda5007/H\beta}) >  \frac{0.61}{\mathrm{\log([\ion{S}{ii}]\lambda 6725/H\alpha) - 0.32}}+1.30
\end{equation}
are classified as AGNs. The [\ion{S}{ii}]$\lambda6725\:\angstrom$ line represents the sum 
of the  [\ion{S}{ii}]$\lambda6717\:\angstrom$ and [\ion{S}{ii}]$\lambda6731\:\angstrom$ lines. 
This procedure resulted
in a sample of 91 objects (35 Sy2 and 56 Sy1) with 
 redshifts $z \: \la \: 0.4$.  
 The reader is referred to Paper~I for a complete description about the observational data  and  aperture effects on metallicity/abundance estimation. 

\begin{figure}
\includegraphics[angle=-90, width=0.47\textwidth]{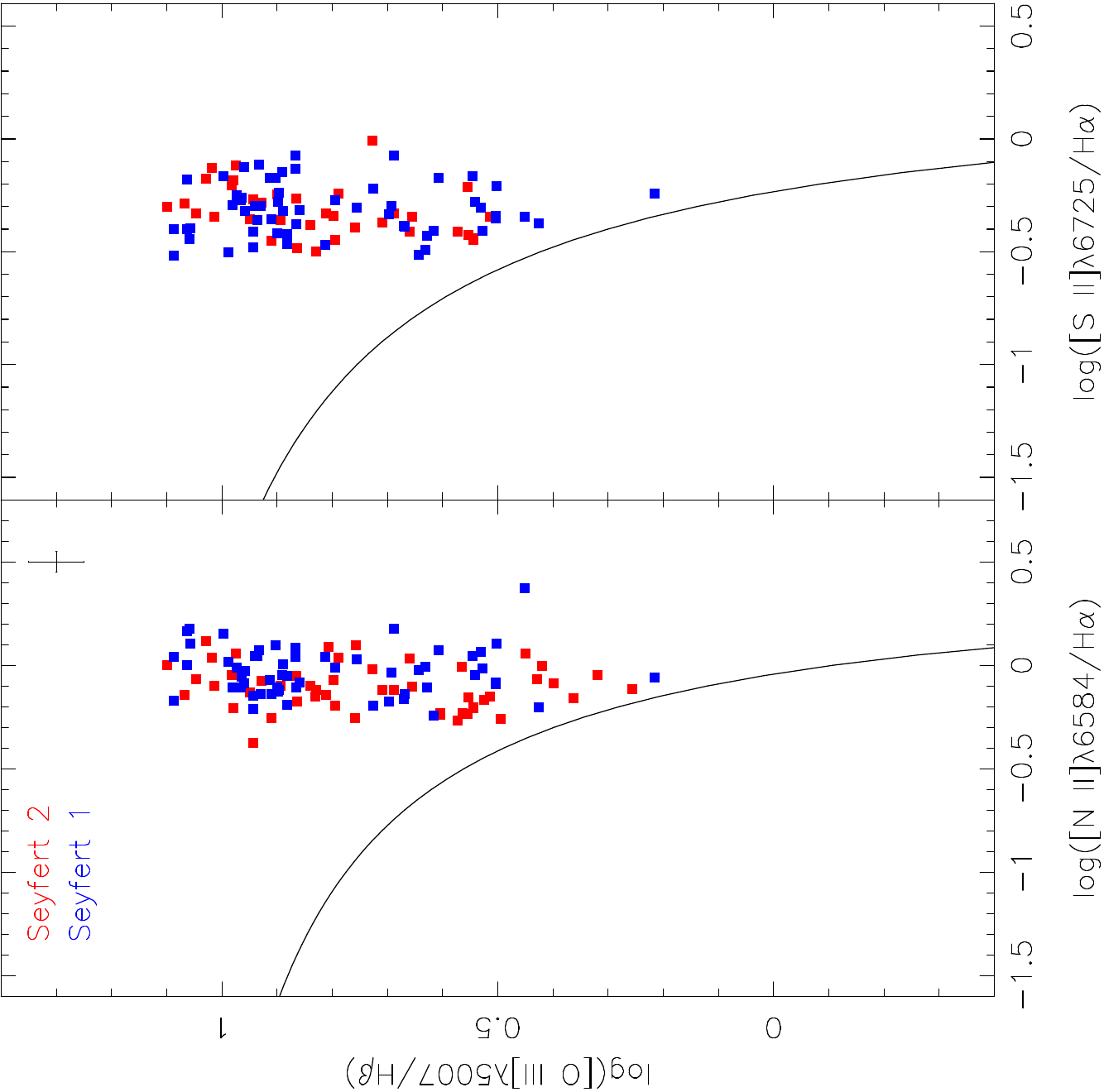}
\caption{Diagnostic diagrams log(\ion{O}{iii}]$\lambda5007$/H$\beta$) versus log[\ion{N}{ii}]$\lambda6584$/H$\alpha$
and versus log([\ion{S}{ii}]$\lambda6725$/H$\alpha$). Points represent objects of our sample (see Sect.~\ref{observ}), where
objects classified as Seyfert 1 and 2 are represented by different colours, 
as indicated. Black lines, taken from  \citet{2001ApJ...556..121K} and represented by Eqs.~\ref{eq1} and \ref{eq2}, separate objects ionized by massive stars   from those ionized by  AGN-like mechanism.
Error bar, in the left panel, represents the typical uncertainty (0.1 dex) in AGN emission-line ratio measurements (e.g., \citealt{kraemer1994spectra}).}
\label{fig1}
\end{figure}

In contrast to previous studies \citep{2020MNRAS.492.5675C, 2020MNRAS.492..468D, 2020MNRAS.496.3209D, 2021MNRAS.501.1370D}, Seyfert~1 nuclei are considered in the present work. According to the unification scheme,  the continuum source and the Broad Line Region (BLR) are blocked by a dusty torus in Seyfert~2 \citep{1993ARA&A..31..473A}. However, it is not clear
if the torus effects extend to  NLRs. For instance, \citet{1998ApJ...506..647S}, 
who compared optical emission line intensities of distinct Seyfert classes, found that Seyfert~1 nuclei
have a higher excitation than Seyfert~2 nuclei. \citet{2008ApJ...685L.109Z} used spectroscopic 
data of AGNs ($z \: < \: 0.3$) from the SDSS DR4 \citep{2017AJ....154...28B} and  
found that Seyferts~1 and 2 have different distributions in the BPT \citep{1981PASP...93....5B}  
[\ion{O}{iii}]$\lambda5007$/H$\beta$ versus
[\ion{N}{ii}]$\lambda6584$/H$\alpha$ diagram.
If Seyferts~1 and 2 of our sample show this difference in diagnostic diagrams some potential biases could be introduced
in the calibrations based on emission line intensities from
both kinds of AGNs. In order to verify this, the diagnostic diagrams [\ion{O}{iii}]$\lambda$5007/H$\beta$ (ordinate)  versus
[\ion{N}{ii}]$\lambda6584$/H$\alpha$ (abscissa) and [\ion{S}{ii}]($\lambda6716$ + $\lambda6731$)/H$\alpha$ (abscissa)  
containing our sample of objects as well as
the criteria to separate SFs from AGNs proposed by \citet{2001ApJ...556..121K},
i.e., Equations~\ref{eq1} and \ref{eq2},  are shown in Figure~\ref{fig1}. In this figure, the line ratios of Seyferts~1 and 2 are indicated by different colours. It can be seen in Fig.~\ref{fig1} that Seyferts 1 and 2 occupy the same regions in both diagrams.
Since the position of an object in diagnostic diagrams is driven by some
physical parameters  of the gas phase (e.g., \citealt{2016MNRAS.456.3354F}), probably,
these objects have similar ionization degree and metallicity. This result is apparently in disagreement with the findings obtained by \citet{2008ApJ...685L.109Z}, however, it is worthwhile to note that these authors considered a sample of objects with wider range of ionization, while 
we selected only objects which have the [\ion{O}{iii}]$\lambda4363\:\angstrom$ line measured, i.e.,
objects with high excitation degree (see also \citealt{2020MNRAS.496.2191F}).

\begin{figure}
\includegraphics[angle=-90, width=0.47\textwidth]{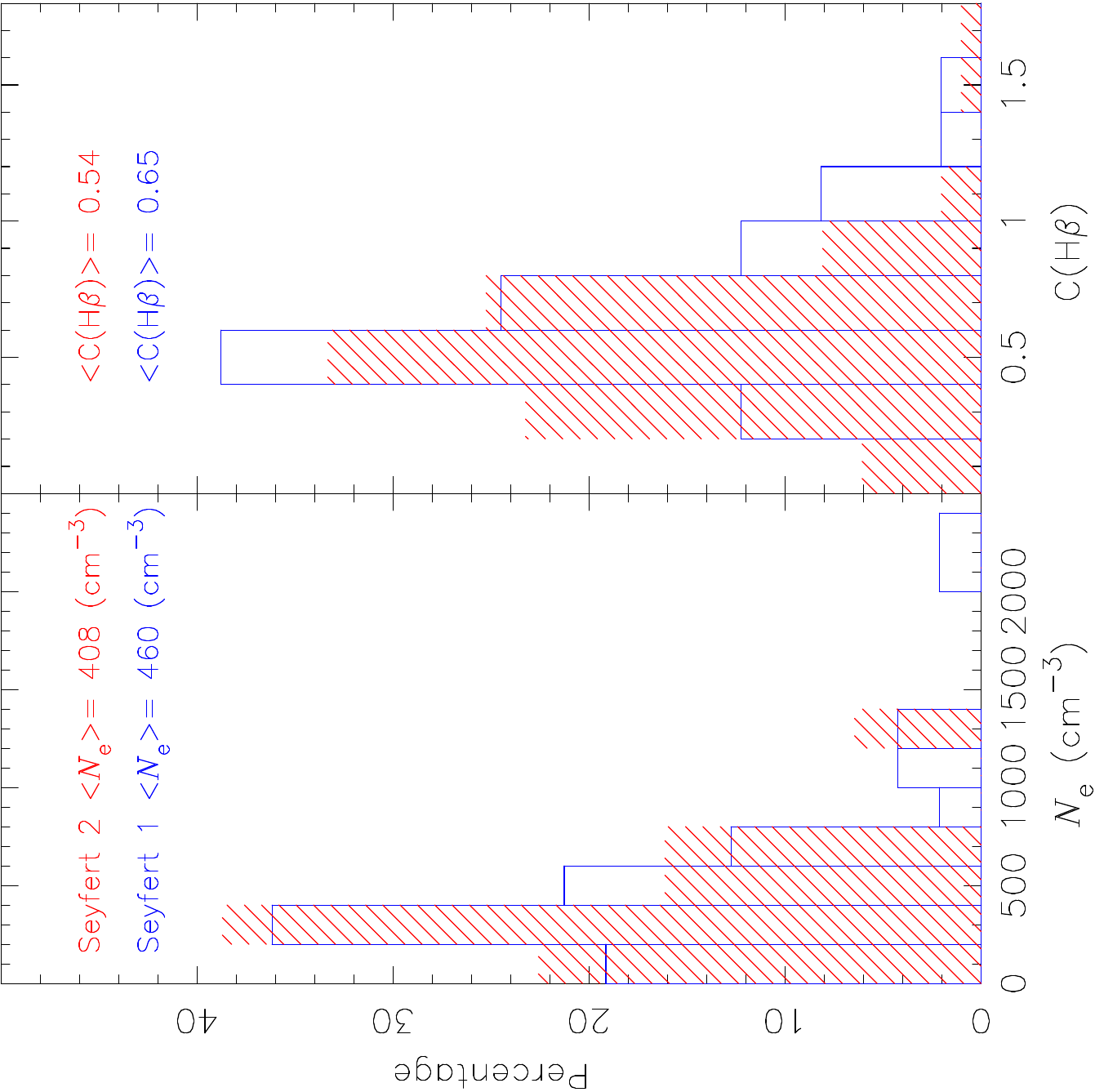}
\caption{Histogram containing the distributions of electron density $N_{\rm e}$ (left panel)
and reddening correction C(H$\beta$) (right panel) for our sample of objects (see Sect.~\ref{observ}).
$N_{\rm e}$ is calculated through the emission line ratio [\ion{S}{ii}]$\lambda6717$/[\ion{S}{ii}]$\lambda6731$.}
\label{fig2}
\end{figure}

Another  concern in our analysis is about the electron density ($N_{\rm e}$) of
Seyfert~1,  because the gas phase in this object class can reach high  $N_{\rm e}$ values
and some lines emitted by transitions between  levels with low critical density $N_{\rm c}$, such as [\ion{O}{ii}]$\lambda3726$ {\AA}, $\lambda3729$ {\AA}  ($N_{\rm c}=10^{3.7}\:\rm  cm^{-3}$, \citealt{2012MNRAS.427.1266V}) can suffer collisional de-excitation, resulting in incorrect abundance estimates. To verify 
the range of  electron density of our sample, the $N_{\rm e}$ value for each object was derived through the relation between this parameter with the  [\ion{S}{ii}]$\lambda6717$/[\ion{S}{ii}]$\lambda6731$ line ratio by using the {\sc iraf} code \citep{1986SPIE..627..733T, 1987JRASC..81..195D, 1995PASP..107..896S} and assuming an electron temperature value of 10\,000 K. In Fig.~\ref{fig2}, left panel,  the $N_{\rm e}$  distribution and the average value for our sample of Seyferts 1 and 2 are shown. It can be seen that both Seyferts types present similar distributions
 and average values of $N_{\rm e}$. The maximum  $N_{\rm e}$ value
 (2250 cm$^{-3}$),   derived for a Seyfert~1 object, is
 a factor of $\sim 2$
 lower than the lowest  critical  density of the lines considered in the
 present analysis. Also in Fig.~\ref{fig2}, right panel, the reddening correction C(H$\beta$) distributions
 and the average values of these  for our sample of objects are shown. We notice very similar
 distributions and average values for both Seyferts~1 and 2. From the analysis above, one can assume that
 the emission lines considered in this work are emitted in 
 the gas phase of Seyferts~1 and 2  with similar physical conditions.  
 
 Typical uncertainty in AGN  emission lines measurements  considered in the present analysis is in order
 of 0.1 dex (e.g., \citealt{kraemer1994spectra}).

\subsection{O/H derivation}
\label{temet}

To calculate the abundance of oxygen in relation to hydrogen (O/H) through the
$T_{\rm e}$-method, we follow the same methodology developed by \citet{2020MNRAS.496.3209D} for AGNs.
 This method is an adaptation of the $T_{\rm e}$-method for \ion{H}{ii} regions
but differing only in the $t_{2}$-$t_{3}$ assumed relation. Thus, hereafter refereed to as
$T_{\rm e}$-method (AGN).

Firstly, using the observational data for each object, the temperature for the  high [$t_{3}(\rm obs.)]$ ionization  gas zone,  i.e., for the nebular region  where $\rm O^{2+}$ and ions with similar ionization
potentials (e.g., $\rm Ne^{2+}$, $\rm S^{2+}$)  are located, is derived from the expression:
\begin{equation}{\label{eqn3.5}}
t_{3}({\rm obs.}) = 0.8254 - 0.0002415\times {R_{\rm O3}} + \frac{47.77}{{R_{\rm O3}}}, 
\end{equation}
\noindent where ${R_{\rm O3}}$ = [\ion{O}{iii}]($\lambda4959$ {\AA} + $\lambda5007$ {\AA})/$\lambda4363$ {\AA} and   $t{_3}$({\rm obs.}) is in units of $10^{4}$ K.  The  $t_{3}({\rm obs.})-R_{\rm O3}$ relation depends on the atomic data used to derive the emissivities of the emission lines involved and it  suffers some uncertainties. In fact, different $t_{3}({\rm obs.})-R_{\rm O3}$ relations have been proposed by several authors along the years (e.g., \citealt{1992MNRAS.255..325P, 2006A&A...448..955I, 2008MNRAS.383..209H, 2014MNRAS.441.2663P}) and different temperature values in order of some hundreds, for a given $R_{\rm O3}$ value, have been derived when distinct relations are considered.
The Eq.~\ref{eqn3.5} is valid for the range ${30\la R_{\rm O3}\la 700}$ with correspondingly temperature range of ${0.70 \: \la \: t_{3}(\rm obs.) \: \la \: 2.3 }$. In this study, only objects with  $t_{3}$(obs.) in this range of values were considered.

Since it is not possible to estimate the temperature for the  low [$t_{2}({\rm obs.}]$) ionization  gas zone,  i.e., for the nebular region where $\rm O^{+}$ and ions with similar ionization potentials (e.g., $\rm S^{+}$, $\rm N^{+}$)  are located, 
due to the absence of the [\ion{N}{ii}]$\lambda$5755  and
[\ion{O}{ii}]$\lambda$7319, $\lambda$7330 observational line intensities,
the following theoretical relation between $t_{2}$-$t_{3}$ has been adopted:
\begin{equation}
\label{t2t3new}
t_{2}=({\rm a} \times t_{3}^{3})+({\rm b} \times t_{3}^{2})+({\rm c} \times t_{3})+{\rm d},
\end{equation}
where $\rm a=0.17$, $\rm b=-1.07$, $\rm c=2.07$, and $\rm d=-0.33$
 and $t{_2}$ is in units of $10^{4}$ K.
 This $t_{2}$-$t_{3}$ relation was derived by \citet{2020MNRAS.496.3209D} based on the photoionization model results built by \citet{2020MNRAS.492.5675C}.
This grid of models takes into account a wide range of nebular parameters, which are summarized  below:
\begin{enumerate}
    \item Spectral Energy Distribution (SED): It is made up of two parts that are added together. The first is a Big Bump component that peaks at $\approx$ 1 Ryd and is parametrized by the temperature of the bump, which is assumed to be $5 \: \times \: 10^{5}$ K. The second component is  an X-ray power law with spectral index $\alpha_x=-1$ which is only added for energies greater than 0.1 Ryd to prevent it from extending into the infrared region. The  $\alpha_{ox}$ spectral index defined as the slope of a power law between 2\: keV and 2500\:\AA\  was assumed to vary from $-0.8$ to $-1.4$
       (see \citealt{2021MNRAS.505.2087K} for a detailed description of this SED). 
\item  Metallicity:  Values for the metallicity in relation to the solar ($Z/{\rm Z_{\odot}}$) = 0.2, 0.5, 0.75, 1.0, 1.5 and 2.0 were assumed in the models.  All the abundances for the heavy metals 
were linearly scaled with the solar abundance\footnote{The solar composition assumed in the
{\sc Cloudy} code is listed in $\href{http://web.physics.ucsb.edu/~phys233/w2014/hazy1_c13.pdf}{http://web.physics.ucsb.edu/~phys233/w2014/hazy1_c13.pdf}$.} with the exception
of nitrogen where we adopted the relation $\rm \log(N/O)=1.29\: \times \: [12+\log(O/H)] \: - \: 11.84$.
\item Electron density ($N_{\rm e}$): The $N_{\rm e}$ was considered to be constant along the radius of AGN 
and the values 100, 500 and 3000 $\rm cm^{-3}$ were assumed.
\item Ionization parameter: The range of the logarithm of the ionization parameter was considered to be $ -4.0 \: \lid \: \log U \: \lid \: -0.5$, with a step of 0.5 dex.
\end{enumerate}
The $t_{2}$ and $t_{3}$ values predicted by the models correspond to the mean temperature for  $\rm O^{+}$ and $\rm O^{2+}$ over the nebular AGN radius  times the electron density.  Therefore, the $t_{2}$(obs.) and $t_{3}$(obs.) temperatures, calculated through  integrated  measurements of the flux of observational emission lines, could differ from those predicted by the models.
In \citet{2021MNRAS.501L..54R}, the  relation $t_{2}$-$t_{3}$ (Eq.~\ref{t2t3new}) was
compared with direct estimations of electron temperatures, calculated from observational
auroral emission lines, for a small sample of AGNs  (11 objects) and a good agreement was found between
them. However, these authors showed that when outflowing gas 
is present in AGNs, a large deviation of direct electron temperature values from
those derived through Eq.~\ref{t2t3new} is obtained.

Additionally, the $\rm O^{2+}/H^{+}$ and $\rm O^{+}/H^{+}$ ionic abundances were estimated using the
following relations: 
\begin{eqnarray}
\label{eqt4}
 12+\log(\frac{{\rm O^{2+}}}{{\rm H^{+}}}) \!\!\!&=&\!\!\! \log \big( \frac{1.33 \times I(5007)}{I{\rm (H\beta)}}\big)+6.144  \nonumber\\
                                          &&\!\!\!+\frac{1.251}{t_{3}({\rm obs.})}-0.55\log t_{3}({\rm obs.}) 
\end{eqnarray}
 and
\begin{eqnarray}
\label{eqt5}
 12+\log(\frac{{\rm O^{+}}}{{\rm H^{+}}})  \!\!\!&=&\!\!\! \log  \big( \frac{I(3727)}{I{\rm (H\beta)}}\big)+5.992 \nonumber\\
                                           &&\!\!\!+\frac{1.583}{t_{2}}-0.681\log t_{2} +\log(1+2.3 n_{\rm e}),\end{eqnarray}
where $n_{\rm e}$ is the electron density $N_{\rm e}$ in  units of 10\,000 $\rm cm^{-3}$.
For each object, the value of $t_{3}({\rm obs.})$ is obtained by using the Eq.~\ref{eqn3.5}. 
 The value of $t_{2}$ is derived by applying $t_{3}({\rm obs.})$ in the theoretical relation represented by Eq.~\ref{t2t3new}. The same procedure is usually carried out in \ion{H}{ii} region abundance studies in scenarios where $t_{2}(\rm obs.$) can not be derived (e.g. \citealt{1992AJ....103.1330G, 2003ApJ...591..801K}.)

Finally, the total oxygen abundance (O/H) was derived assuming
\begin{equation}
\label{eqt6}
{\rm
\frac{O}{H}=ICF(O)\: \times \: \left[\frac{O^{2+}}{H^{+}}+\frac{O^{+}}{H^{+}}\right],} 
\end{equation}
where  ICF(O) is the Ionization Correction Factor for oxygen which takes into account the contribution of  unobserved oxygen ions (e.g., $\rm O^{3+})$. To derive ICF(O) it is necessary to have
the $\rm He^{+}/H^{+}$ and  $\rm He^{2+}/H^{+}$ abundances (e.g., \citealt{1977RMxAA...2..181T, 2006A&A...448..955I}), which it is not possible to
derive because the helium recombination line  $\lambda 4686\:\angstrom$ is not available in our data sample. Therefore, we assume for all objects an average 
value of 1.20 for ICF(O), which translates into an abundance correction
of 0.1 dex, i.e., in order of the uncertainty derived in $T_{\rm e}$-method estimates
(e.g., \citealt{2003ApJ...591..801K, 2008MNRAS.383..209H}).
This value represents the average for the values derived by  \citet{2020MNRAS.492..468D}, who found ICF(O) values ranging from 1.00 to 1.80 for a sample of local Seyfert~2.

We assume the typical uncertainty in the O/H estimates to be in the order of 0.1 dex, as estimated
in \citet{2020MNRAS.496.2191F}.

\section{O/H calibration}
\label{res}

To calibrate a certain  line ratio with the abundance, initially, it is necessary to analyse if the line ratio being considered has a secondary dependence on   
other physical parameters, usually, on the ionization degree of the gas. In the case of the $R_{23}$ line ratio,
\citet{1991ApJ...380..140M} suggested that its use as O/H indicator for SFs must be conciliated with the $O32$=([\ion{O}{iii}]$\lambda4959+\lambda5007$)/[\ion{O}{ii}]$\lambda3727$,
where [\ion{O}{ii}]$\lambda3727$ is the sum of $\lambda3726$ and $\lambda3729$.
The $O32$ line ratio has a strong dependence on the ionization degree of the gas and/or
on the effective temperature of the hottest ionizing stars of SFs (see \citealt{2001A&A...369..594P, 2003A&A...404..969D, 2017MNRAS.466..726D}). Thus, any $R_{23}$ calibration would consider line ratios dependent on the hardness of the ionizing radiation \citep{2001A&A...369..594P}.

\begin{figure}
\includegraphics[angle=0, width=0.47\textwidth]{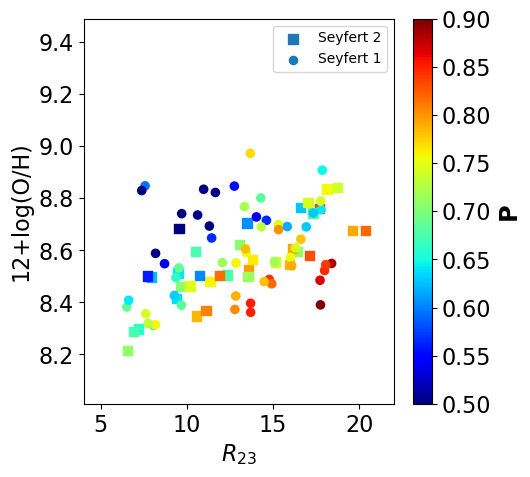}
\caption{Oxygen abundance [in units of 12+log(O/H)] versus
$R_{23}$ for our sample of objects (see Sect.~\ref{observ}).
O/H abundances are calculated by using the  $T_{\rm e}$-method (AGN). Seyferts~1 and 2 nuclei are represented by different symbols
and colour bars indicate the P value for each object.}
\label{fig3}
\end{figure}

In relation to the $R_{23}$ application as O/H abundance indicator for AGNs, \citet{dors2015central}, by using results of a grid of photoinization models,
showed that the $R_{23}$-O/H  relation is dependent on the number of ionizing photos $Q(\rm H)$ or on the ionization parameter ($U$) of the gas [$U \varpropto Q(\rm H)$]. Following \citet{2001A&A...369..594P}, 
we adopt the line ratio defined as
\begin{equation}
{\rm P}=[([\ion{O}{iii}]\lambda4959+\lambda5007)/{\rm H}\beta]/R_{23}    
\end{equation}
 as an indicator 
of the hardness of the ionizing radiation.
In order to verify the dependence of $R_{23}$ on the hardness of the ionizing radiation
by using our data, in Fig.~\ref{fig3}, the oxygen abundance  derived through the
$T_{\rm e}$-method (AGN) for each object of our data sample versus the corresponding  $R_{23}$ value is shown.
 In this figure, the colour bars indicate objects with different P values while objects classified as Seyferts~1 and 2 are represented by different symbols. We notice the following:
\begin{enumerate}
    \item  Seyferts~1 and 2 show similar O/H abundances and P values, and
     
    \item the  O/H-$R_{23}$ relation for AGNs is dependent on P, hence objects with lower P values are located at the top-left region in Fig.~\ref{fig3}.
\end{enumerate}

\begin{figure}
\includegraphics[angle=0.0, width=0.52\textwidth]{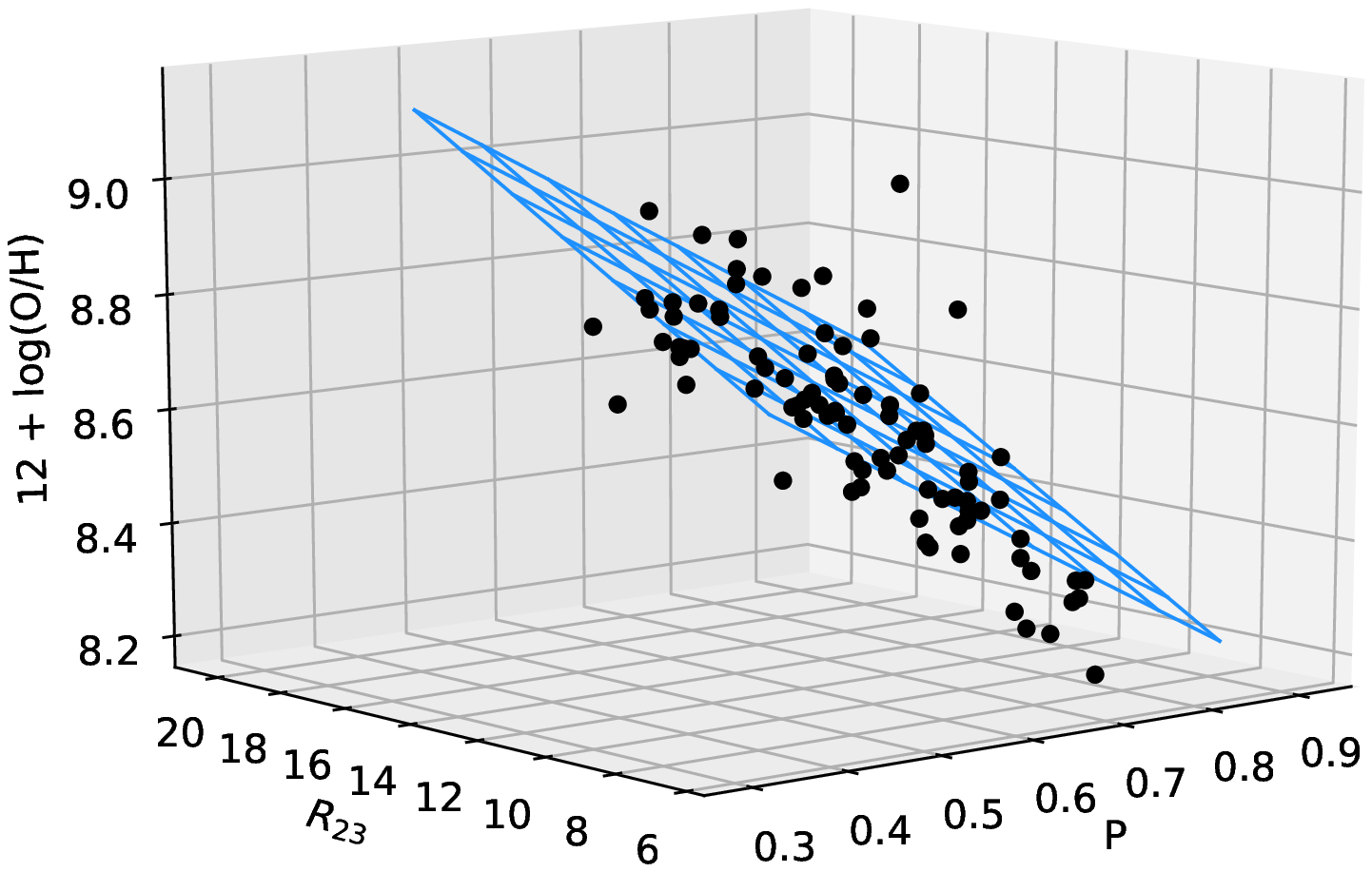}
\caption{Bi-dimensional calibration between the oxygen abundance
[in units of 12+log(O/H)], $R_{23}$=([\ion{O}{ii}]$\lambda3726+\lambda3729$ + [\ion{O}{iii}]$\lambda4959+\lambda5007$)/H$\beta$ and P=(\ion{O}{iii}]$\lambda4959+\lambda5007$/H$\beta$)/$R_{23}$.
The surface represents the   fit to the points  given by Eq.~\ref{eq8}.
Points represent the line ratio intensities of our sample of objects
(see Sect.~\ref{observ}) with the corresponding 
 O/H abundance   obtained by using the  $T_{\rm e}$-method (AGN)
(see Sect.~\ref{temet}).}
\label{fig5}
\end{figure}

The fact that O/H-$R_{23}$ relation is dependent on P indicates that a bi-parametric 
calibration O/H=f($R_{23}$, P) is more accurate for AGNs instead of O/H=f($R_{23}$). In view of this, the $R_{23}$, P and 12+log(O/H) values are shown in 
Fig.~\ref{fig5}. A fit to the points,  by using the least square method, results in the following expression 

\begin{equation}
\label{eq8}
Z = (-1.00 \pm  0.09){\rm P} + (0.036 \pm 0.003)R_{23} + (8.80 \pm 0.06),   
\end{equation}
where $Z\equiv12+\log(\rm O/H)$. We refer to this approach as the D-method.

\section{Discussion}
\label{disc}

The estimation of O/H abundance through strong emission lines was first proposed by 
\citet{1976ApJ...209..748J}. Based on the behaviour of the intensity of optical emission line ratios
across the disk of some nearby spiral galaxies, mainly caused by radial abundance gradients
as originally proposed by \citet{1971ApJ...168..327S}, \citet{1976ApJ...209..748J} suggested that the  
[\ion{O}{iii}]$\lambda5007$/H$\beta$ and [\ion{N}{ii}]$\lambda6584$/[\ion{O}{ii}]$\lambda3727$
line ratios can be chemical enrichment indicators. Between these two line ratios,
these authors posited that [\ion{O}{iii}]/H$\beta$ shows some advantages 
over [\ion{N}{ii}]/[\ion{O}{ii}], however, any calibration between these line ratios
and O/H can be obtained. Thereafter, \citet{10.1093/mnras/189.1.95} introduced
the $R_{23}$ line ratio as an O/H abundance indicator. These authors, by using O/H
estimates based on $T_{\rm e}$-method
for the disk of \ion{H}{ii} regions and predictions from  theoretical models built by  \citet{1976ApJ...208..323S}, \citet{1978ApJ...222..821S} and \citet{1978A&AS...32..429S}, proposed the first
calibration between strong line ratios and O/H abundance for SFs (for  other pioneering papers see, for instance, \citealt{1979A&A....78..200A,  1981A&A....93..362S, 1983MNRAS.204...53S, 1984MNRAS.211..507E}). 
The first empirical calibration between strong emission lines and O/H abundances derived through the $T_{\rm e}$-method  seems to have been proposed by \citet{1994ApJ...429..572S}\footnote{For a review on empirical calibration for SFs and their limitations see \citet{2019ARA&A..57..511K}.}, who presented a SF calibration  using the
$N2$=([\ion{N}{ii}]$\lambda6548+\lambda6584$)/H$\alpha$  line ratio as abundance indicator (see also
\citealt{2004MNRAS.348L..59P,  2007A&A...475..409S, 2007A&A...462..535Y, 2007A&A...473..411L, 2012MNRAS.424.2316P, 2013A&A...559A.114M, 2014ApJ...797...81M, 2015ApJ...813..126J,  2016ApJ...825L..23S,  2016MNRAS.458.1529B, 2016MNRAS.457.3678P, 2018ApJ...859..175B, 2019ApJ...872..145J, 2019ApJ...887..168G}).

In regard to AGNs, the first strong emission line calibrations were proposed by \citet{1998AJ....115..909S}, who by using photoinization model
results, proposed two bi-dimensional calibrations among 
$N2$-[\ion{O}{iii}]/H$\beta$, $N2$-[\ion{O}{ii}]/[\ion{O}{iii}] narrow
line ratios and  O/H abundance. Thereafter, theoretical (e.g., \citealt{2014MNRAS.443.1291D, dors2015central}) and semi-empirical (e.g., \citealt{2017MNRAS.467.1507C, 2020MNRAS.492.5675C, 2021MNRAS.501.1370D})
calibrations  as well as bayesian-like approach to derive AGNs chemical abundances (e.g., \citealt{2018ApJ...856...89T, 2019A&A...626A...9M, 2019MNRAS.489.2652P}) have been proposed.
Recently, \citet{2020MNRAS.496.2191F} developed an approach for estimating abundances of heavy elements, which
involves a reverse-engineering of the $T_{\rm e}$-method and considering
the [\ion{O}{iii}]$\lambda$5007/H$\beta$ versus [\ion{N}{ii}]$\lambda$6584/H$\alpha$ diagnostic diagram.
This methodology, which is only based on strong emission lines, consistently recovers O/H and N/H abundance values
calculated through the $T_{\rm e}$-method with an uncertainty of about 0.2 dex. The method proposed by \citet{2020MNRAS.496.2191F}, although has been classified by them as semi-empirical  calibration,
it is the first AGN calibration which takes into account O/H  as reference values derived
through the $T_{\rm e}$-method instead of photoionization models.

 In the present work we proposed an empirical calibration for AGNs
between $R_{23}$ and P line ratios with the O/H abundance, which is represented by a simple bi-dimensional expression (Eq.~\ref{eq8}). 
In what follows this approach (D-method) is a somewhat thoroughly discussions of its implications relative to other known methods in AGNs studies.
First of all, there are some concerns on the abundance determinations in AGNs via $T_{\rm e}$-method.
For instance, \citet{1984A&A...135..341S} compared observational emission line ratio intensities
of a sample of Seyfert~2 AGNs with those predicted by photoionization models. 
 This author argued that the [\ion{O}{iii}]($\lambda4959+\lambda5003$)/$\lambda$4363 ratio
 is enhanced by gas emission with very high electron density which precludes any
 abundance estimations in AGNs via $T_{\rm e}$-method (see also \citealt{2001ApJ...549..155N}).
 In fact, direct estimates of O/H abundance  based on the same methodology of the  $T_{\rm e}$-method
 for SFs applied to AGN studies produces unreal low (subsolar) O/H values (see Figure 6 of  Paper~I) for most part of objects. However, these
 low values are not due to electron density effects but as a consequence of inappropriate use of the relations between temperatures of the low  and high  ionization
gas zones derived for SFs in AGN chemical abundance studies (see \citealt{2020MNRAS.496.3209D, 2021MNRAS.501L..54R}).
Therefore, our O/H estimates following the adaptation of the $T_{\rm e}$-method
for AGNs by \citet{2020MNRAS.496.3209D}, in principle, are correct and can be
used to derive empirical calibrations based on strong emission lines of AGNs.
In fact, by inspecting Fig.~\ref{fig3} carefully, it can be seen that  most parts of the sample present
 O/H abundance values close to or over the solar abundance [12+log(O/H)$_{\odot}=8.69$, \citealt{allendeprieto01}]
 and few objects present equally subsolar  O/H values (see also \citealt{2006MNRAS.371.1559G}).

\begin{figure}
\includegraphics[angle=-90.0, width=0.47\textwidth]{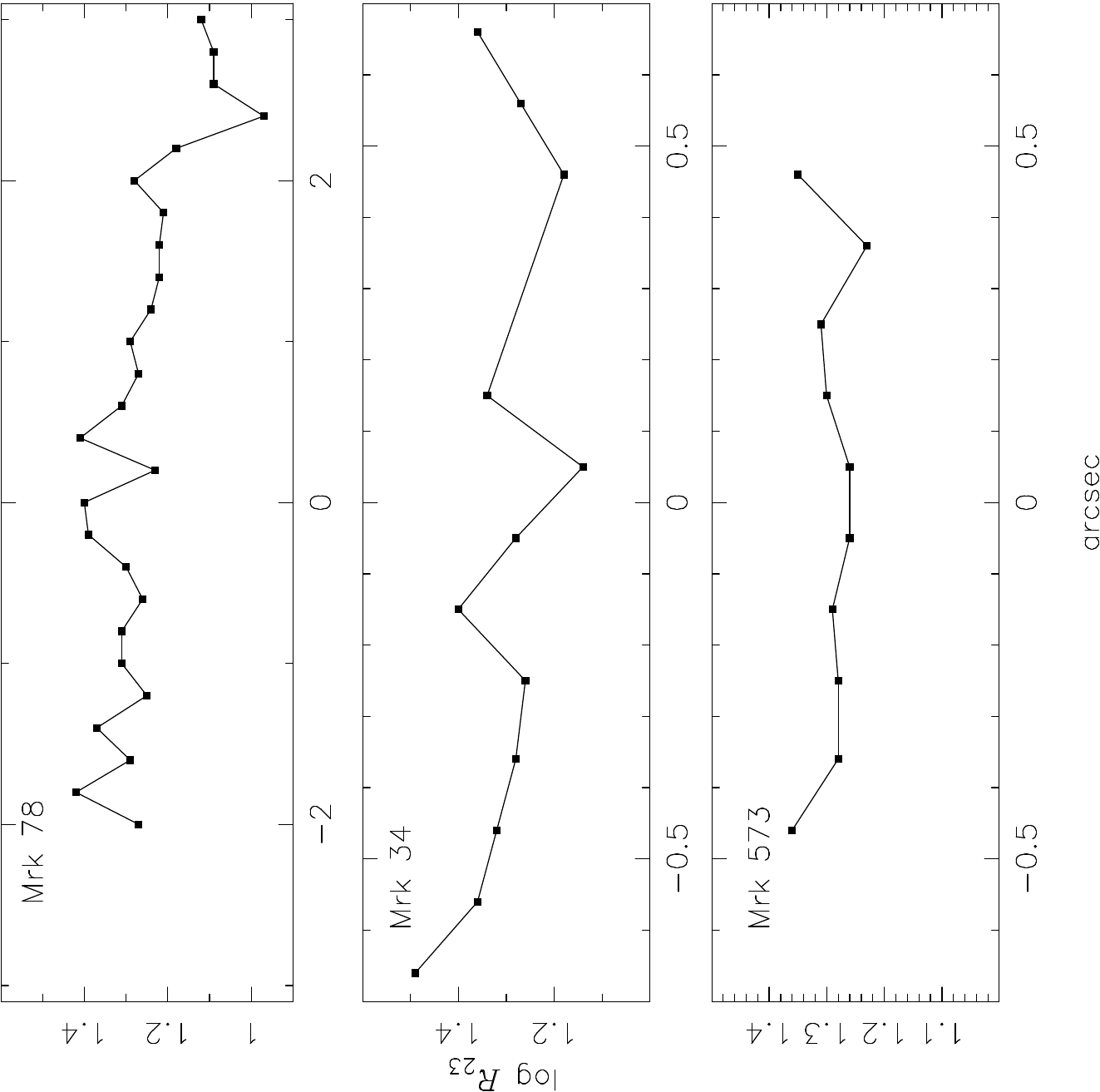}
\caption{Logarithm of $R_{23}$ as a function of distance (in arcsec) from the centre 
of three AGNs, as indicated. The data for Mrk~573, Mrk~34 and Mrk~78
were taken from \citet{2018ApJ...856...46R}, \citet{2018ApJ...867...88R}
and \citet{2021arXiv210106270R}, respectively.
For Mrk~573, Mrk~34 and Mrk~78  1 arcsec corresponds 
to  349.1, 1007.7 and 747.4 pc in the galactic plane of each object, respectively
\citep{2021arXiv210106270R}.}
\label{fig6}
\end{figure}

 Another concern in that  the value of an abundance indicator must not differ substantially across the area of the object type for which it will be considered in the estimation of abundances. Otherwise, 
the value of the abundance indicator measured from integrated flux might not be representative of the entire object.
For SFs, \citet{2000ApJS..128..511O} found no spatial variation of the 
$R_{23}$ across \ion{H}{ii} regions with distinct morphology (see also \citealt{2000ApJ...539..687O, 2001A&A...369..594P, 2010MNRAS.402.1635R, 2018ApJ...853..151M}). In order to verify this in our case, the logarithm of $R_{23}$ as a function of  the distance (in arcsec) from the centre 
of three AGN nuclei Mrk\,573, Mrk\,34 and Mrk\,78, whose data were taken
from  \citet{2018ApJ...856...46R}, \citet{2018ApJ...867...88R}
and \citet{2021arXiv210106270R}, respectively, is shown in Fig.~\ref{fig6}.
Although a  small decreasing of $R_{23}$ 
is noted at external region (at distances larger than 2 arcsec or $\sim 1.5$ kpc)
from the centre of Mrk\,78, we notice that  $R_{23}$ is relatively constant
within the three objects. Therefore, similar to \ion{H}{ii} regions,
$R_{23}$ is a robust oxygen abundance indicator for AGNs.

\begin{figure}
\includegraphics[angle=-90.0, width=0.47\textwidth]{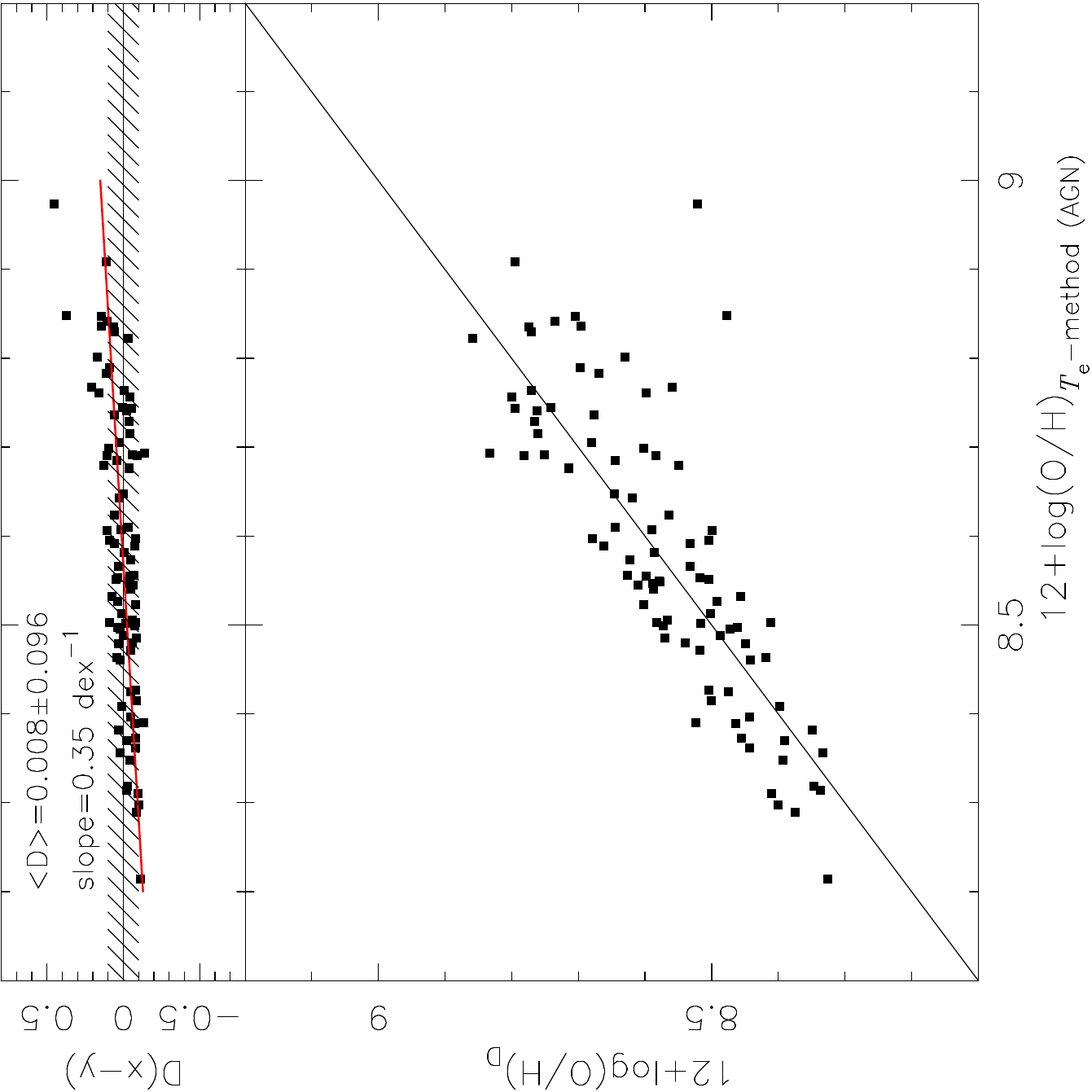}
\caption{ Bottom panel: Comparison between oxygen abundances [in units of 12 + log(O/H)] computed using the observational data described in Sect.~\ref{observ}  obtained through the  D-method (y-axis, Eq.~\ref{eq8}) and $T_{\rm e}$-method (AGN) (x-axis, see Sect.~\ref{temet}).
Solid line represents the equality between the estimates.  Top panel: 
Difference (D = x–y) between the estimations versus those via $T_{\rm e}$-method. 
The dashed area indicates the uncertainty of $\pm 0.1$ assumed in the oxygen abundance
estimations (e.g., \citealt{2002MNRAS.330...69D}), whilst red line represents a linear regression to these differences whose
slope is indicated. The average difference (< D >) is indicated.}
\label{fig7}
\end{figure}

In Fig.~\ref{fig7}, bottom panel, a comparison between oxygen abundance [in units
of 12+log(O/H)] derived by using the D-method (Eq.~\ref{eq8}) 
with estimations from the $T_{\rm e}$-method (AGN)  is performed.
In Fig.~\ref{fig7}, top panel the difference (D) between these two methods 
versus the $T_{\rm e}$-method (AGN) estimates is shown. 
There is a good agreement between the estimates for a wide range of O/H abundances, with D being approximately zero.
However, we notice that the D-method (over) underestimates the O/H for the very (low) high  metallicity regimes. In spite of the above observations, more objects
with low and high metallicities are necessary to obtain a better conclusion.
A comparison between O/H estimates derived assuming a vast number
of methods available in the literature is presented in Paper~I and it is repeated here.

\begin{figure}
\includegraphics[angle=-90.0, width=0.47\textwidth]{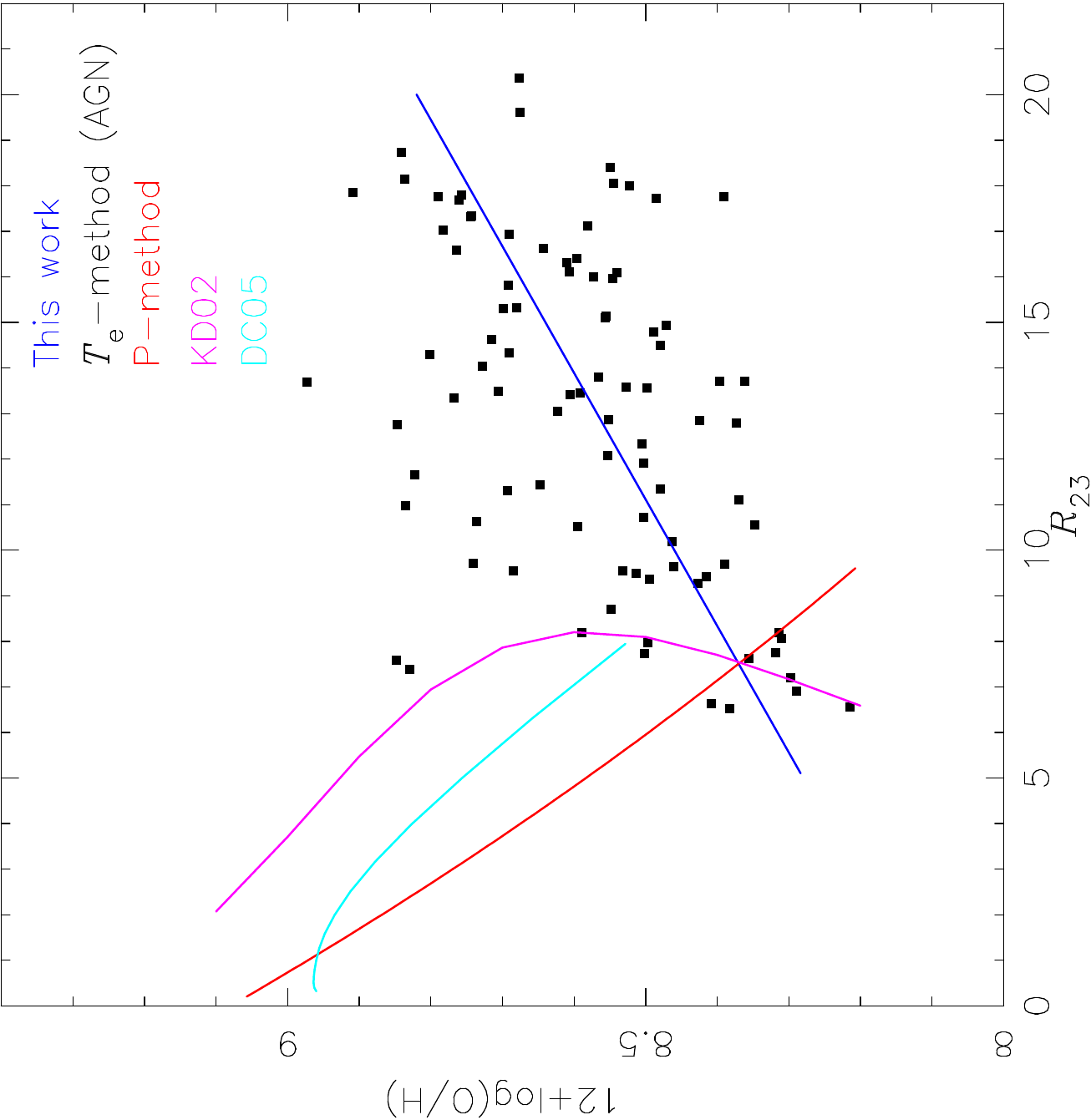}
\caption{Oxygen abundances [in units of 12 + log(O/H)] versus $R_{23}$.
Red, cyan and pink lines represent SF calibrations proposed by 
\citet{2001A&A...369..594P} (P-method, Eq.~\ref{OHR23f}), \citet{2005A&A...437..837D} (DC05) and \citet{2002ApJS..142...35K} (KD02), respectively, as indicated. Blue line represents our calibration (Eq.~\ref{eq8}).
Points represent estimates via  $T_{\rm e}$-method (AGN) for our sample.
Our calibration and the one by \citet{2001A&A...369..594P}  are shown for P=0.7.}
\label{fig8}
\end{figure}

Finally, we proceed with a comparison between our AGN  O/H calibration and that of SFs. Particularly, this comparison can provide  important pieces information on the SED,
 physical processes in the gas phase of AGNs  and
how  they differ from those in SFs.
In the light of the foregoing, the O/H-$R_{23}$, P relation for AGNs (Eq.~\ref{eq8}) and the empirical calibration
\begin{equation}
12+\log({\rm O/H})_{\rm P} = \frac{R_{23} {\rm + 54.2  + 59.45 P + 7.31 P^{2}}}
                       {6.07  + {\rm 6.71 P + 0.371 P^{2} +} 0.243 R_{23}}
\label{OHR23f}
\end{equation}
for O/H determinations in SFs with moderately high metallicity [$\rm 12+\log(O/H) \: \ga \: 8.2$] proposed by \citet{2001A&A...369..594P} are shown in Fig.~\ref{fig8}. In both calibrations
a fixed value of  P=0.70 is assumed,  which is the mean value for our sample of AGNs. Also in Fig.~\ref{fig8}, the $R_{23}$ and O/H  values  derived via $T_{\rm e}$-method (AGN) for our sample of objects, the semi-empirical   calibration derived 
by \citet{2005A&A...437..837D} and the theoretical calibration (assuming the ionizing parameter $q=4 \:\times \: 10^{7} \: \rm cm \: s^{-1}$) proposed by  \citet{2002ApJS..142...35K} are shown. This $q$ value represents an average value from the range of values considered
by \citet{2002ApJS..142...35K}. It can be seen from Fig.~\ref{fig8} that our calibration
and the $T_{\rm e}$-method (AGN) estimates have a different behaviour between O/H-$R_{23}$
relative to the calibrations for SFs, in the
sense that $R_{23}$ increases with the increase in O/H. In SFs, for the metallicity
regime considered in Fig.~\ref{fig8}, the decrease in $R_{23}$  with the increase in O/H  is mainly due  to the decrease in the electron temperature and  electrons with enough energy to excite the levels involved in producing the optical collisional excitation lines (CELs), in the case, $R_{23}$.  
Moreover, as increase in the  metallicity  occur,  
effects of line blanketing by metal lines in the  atmosphere of the ionizing star(s) become more pronounced (e.g., \citealt{2013ApJ...769...94Z}),  resulting  in  softer SED in stars and, consequently, decreasing  $R_{23}$.
 A decrease  in the ionization parameters ($U$) with the
increase of  metallicity (or O/H) can also contribute to a decrease in $R_{23}$. However, the dependence relations between these parameters is controversial in the literature
 (see \citealt{2019MNRAS.483.1901Z} and references therein).

For AGNs, probably,  SEDs are not affected by line blanketing or the NLRs are not
significantly modified by changes in the SED,  producing a direct relation between O/H and $R_{23}$.  \citet{2012ApJ...756...51L} 
presented spectroscopic observations of 27  AGNs with some of the lowest black hole (BH)
masses known and compared the emission line ratios and  SEDs of these objects with that
of AGNs with higher-mass BHs. These authors found evidence for steeper far-UV spectral slopes in lower-mass systems
in comparison with higher mass BHs. However, \citet{2012ApJ...756...51L} found similar
NLR emission lines in objects with distinct BH masses. Also, \citet{2013MNRAS.431..836S} 
showed that parameters other than the ionizing continuum slope, such as metallicity, density and ionization parameter, dominate the scatter in the BPT diagrams.
 Regarding the $U$-O/H dependence for AGNs, 
\citet{2019MNRAS.489.2652P}, based on  bayesian comparison between optical emission-line intensity ratios
of a sample of AGNs and photoionization model results, found no correlation between these parameters. However, the sample considered by these authors consists
of few objects (47 Seyfert 2), therefore, additional analysis is necessary to confirm this result.

\section{Conclusions}
\label{conc}

The oxygen abundance estimates based on strong emission lines of Seyfert nuclei were investigated.
Starting from the idea that the [\ion{O}{ii}]$\lambda\lambda$3726, 3729 and [\ion{O}{iii}]$\lambda$5007
emission lines contain the necessary information to estimate the oxygen abundance in Seyfert nuclei
and the $T_{\rm e}$-method,  adapted for AGNs, yields reliable abundance values, an empirical abundance calibration
among  $R_{23}$, P and O/H was derived for AGNs. This calibration has been derived  by using intensities of emission line ratios from a sample of Seyferts~1 and 2 nuclei, whose data were 
taken from the SDSS DR7 and the lines measurements carried out by the  MPA/JHU group.
No variation of $R_{23}$ is observed inside AGNs, showing that this line ratio is a good O/H abundance indicator.
The derived O/H=f($R_{23}$, P) relation produces, for a wide range of 
metallicities [$\rm 8.0 \: < \: 12+\log(O/H) \: < \: 9.2$],
oxygen abundance values similar to those estimated via $T_{\rm e}$-method (AGN).  In contrast to star-forming regions,
the $R_{23}$ line ratio increases with the increase of O/H in AGNs, indicating that the hardness of ionizing 
radiation is not affected by the metallicities in these kinds of objects or NLRs 
are not significantly  modified by changes in the SED due to metallicity variations.

\section*{Acknowledgements}

 OLD is grateful to Fundacão de Amparo à
Pesquisa do Estado de São Paulo (FAPESP) and Conselho Nacional
de Desenvolvimento Científico e Tecnológico (CNPq).
\section*{Data Availability}

The data underlying this article will be shared on reasonable request
to the corresponding author.

\bibliographystyle{mnras}
\bibliography{paper} 







\bsp	
\label{lastpage}
\end{document}